\documentclass[a4paper,11pt]{article}
\usepackage{pos}
\usepackage{lineno}

\title{Piritakua: the atmosphere as a high-energy physics laboratory}

\author*[a]{Hermes Le\'on Vargas}
\author[a]{Antonio Galv\'an}
\author[a]{Andr\'es Sandoval}
\author[a]{Ernesto Belmont}
\author[a]{Cindy Castell\'on Salguero}
\author[b]{Adiv Gonz\'alez Mu\~noz}

\affiliation[a]{Instituto de F\'isica UNAM,\\
  Circuito de la Investigaci\'on Cient\'ifica, Ciudad Universitaria, CDMX, M\'exico.}

\affiliation[b]{TecNM/Instituto Tecnol\'ogico de Oaxaca,\\
Av. Ing. V\'ictor Bravo Ahuja, Oaxaca, M\'exico.}

\emailAdd{hleonvar@fisica.unam.mx}
\emailAdd{edwin@fisica.unam.mx}
\emailAdd{asandoval@fisica.unam.mx}
\emailAdd{belmont@fisica.unam.mx}
\emailAdd{cindym@estudiantes.fisica.unam.mx}
\emailAdd{adiv.gonzalez@itoaxaca.edu.mx}

\abstract{The atmosphere provides a large set of experimental conditions on which cosmic-ray induced high-energy hadron interactions can take place. These conditions include: sudden changes in the atmospheric pressure, temperature, and in the local electric and magnetic fields. In this talk we introduce the Piritakua (flash of lightning, in the language of the pre-Columbian Pur\'epecha Empire in Mexico) project, a cosmic-ray detector located at the Instituto de F\'isica of UNAM, in Mexico City at 2280 m. a.s.l. The experiment consists of a small array of scintillator detectors, which use the electronics developed by the CosmicWatch project. The scintillators operate simultaneously with an electric field meter, a magnetometer, a meteorological station, and a hemispheric camera. We propose to use Piritakua to study the modification of the secondary particle production and propagation under sudden variations in the standard atmospheric properties. We present the current status and the first results of the experiment.}

\ConferenceLogo{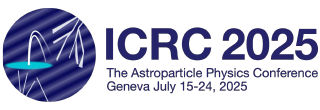}

\FullConference{39th International Cosmic Ray Conference (ICRC2025)\\
 15–24 July 2025\\
Geneva, Switzerland\\}

\begin{document}
\maketitle

\section{The Piritakua project and high-energy atmospheric physics}

Cosmic ray experiments that detect the secondary particles produced by air showers inevitably measure periodic and aperiodic intensity variations in the particle flux. These are mainly produced by changes in the atmospheric pressure, temperature, air density and humidity, that induce changes in the column density above the particle detectors. Broadly speaking, changes in the density of the atmosphere produce changes in the lateral spread of particles in the air showers due to scatterings, and an increase in mass above the detector will enhance particle absorption. Moreover, the electromagnetic properties of the atmosphere accelerate and deflect the secondary charged particles. Therefore, in order to perform precision studies using air showers, it is crucial to understand these complex set of systematic effects.

The primary goal of the Piritakua project is to characterize the effect that transient changes in the atmospheric properties have on the production and propagation of secondary particles from air showers. A secondary goal is to study if it is possible to measure hard radiation phenomena from thunderclouds with a small surface detector. These phenomena include: gamma-ray glows (also called Thunderstorm Ground Enhancements, TGEs), that happen at time scales that go from 1 to 100 seconds, terrestrial gamma-ray flashes (TGFs), that last few hundred microseconds and are accompanied by radio and optical signals, and the recently discovered new phenomenon \cite{flickering}, called Flickering Gamma-ray Flashes, that happen at time scales between those of gamma-ray glows and TGFs, but that are radio and optically silent. This last phenomenon has been proposed to be related to lightning initiation. 

Our experiment is located on the main campus of the National Autonomous University of Mexico (UNAM) in Mexico City (2280 m. a.s.l., $19^{\circ}19'2''$N, $99^{\circ}10'37''$W). The altitude above sea level of Mexico City makes it an interesting site to study phenomena related to the atmospheric electric fields. For instance, the electric field strength required to initiate a relativistic runaway electron avalanche decreases with increasing altitude \cite{lhasso}, due to the reduction of the air density. The critical field at the elevation of Mexico City is expected to be $\approx$ 24$\%$ smaller than at sea level. 

\begin{figure}[h]
\centering
\begin{minipage}{.45\linewidth}
  \includegraphics[width=\linewidth]{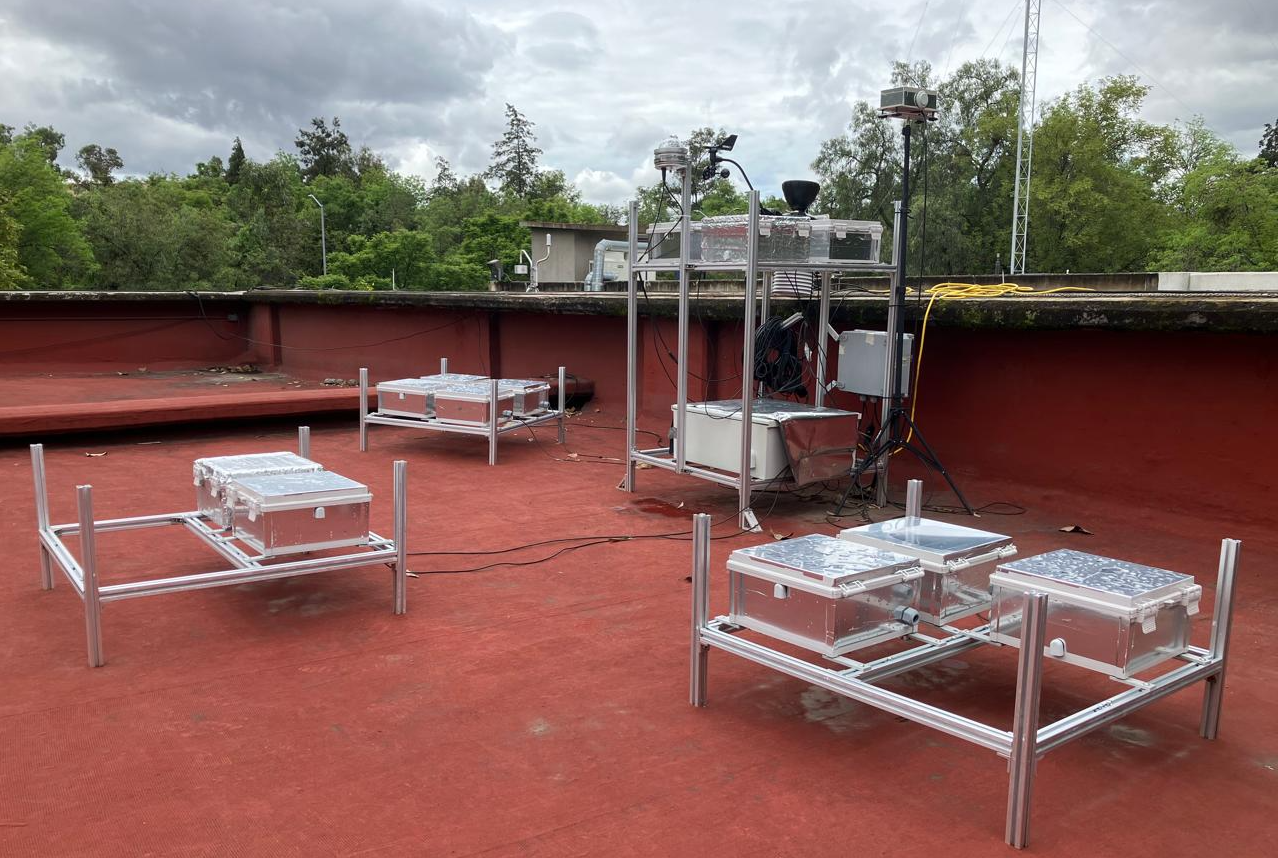}
  \caption{Current construction status of the Piritakua project. \hfill \newline \hfill} 
  \label{fig-1}
\end{minipage}
\hspace{.05\linewidth}
\begin{minipage}{.45\linewidth}
  \includegraphics[width=\linewidth]{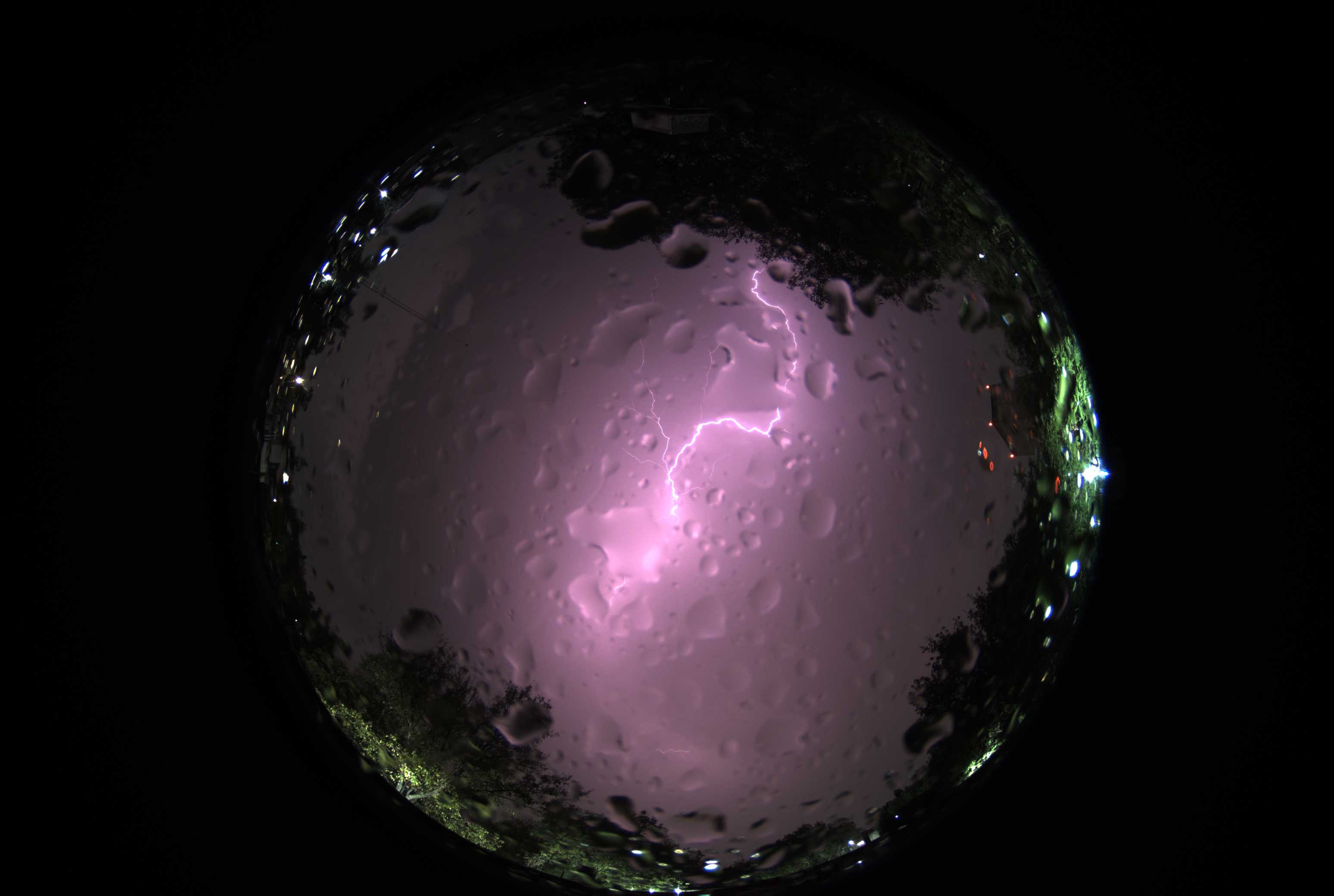}
  \caption{Photograph of lightning above the location of the experiment, during a storm on June 11, 2025.}
  \label{fig-2}
\end{minipage}
\end{figure}

\section{The array of sensors of Piritakua}

Our experiment is currently under commissioning. A picture of its current status can be seen in Figure \ref{fig-1}. The experiment is located at the rooftop of the Experimental Physics Department of UNAM's Physics Institute. It comprises a Davis Vantage Pro2 weather station, an ALPHEA "All Sky" camera, an electric field ERL-10 dual sensor from Boltek and a three-axis Sensys fluxgate magnetometer. The electric field dual sensor is composed of an electric field mill with a maximum range of approximately 40 km, and a lightning detector with a maximum range of approximately 480 km. 

\begin{figure}[h]
\centering
\begin{minipage}{.45\linewidth}
  \includegraphics[width=\linewidth]{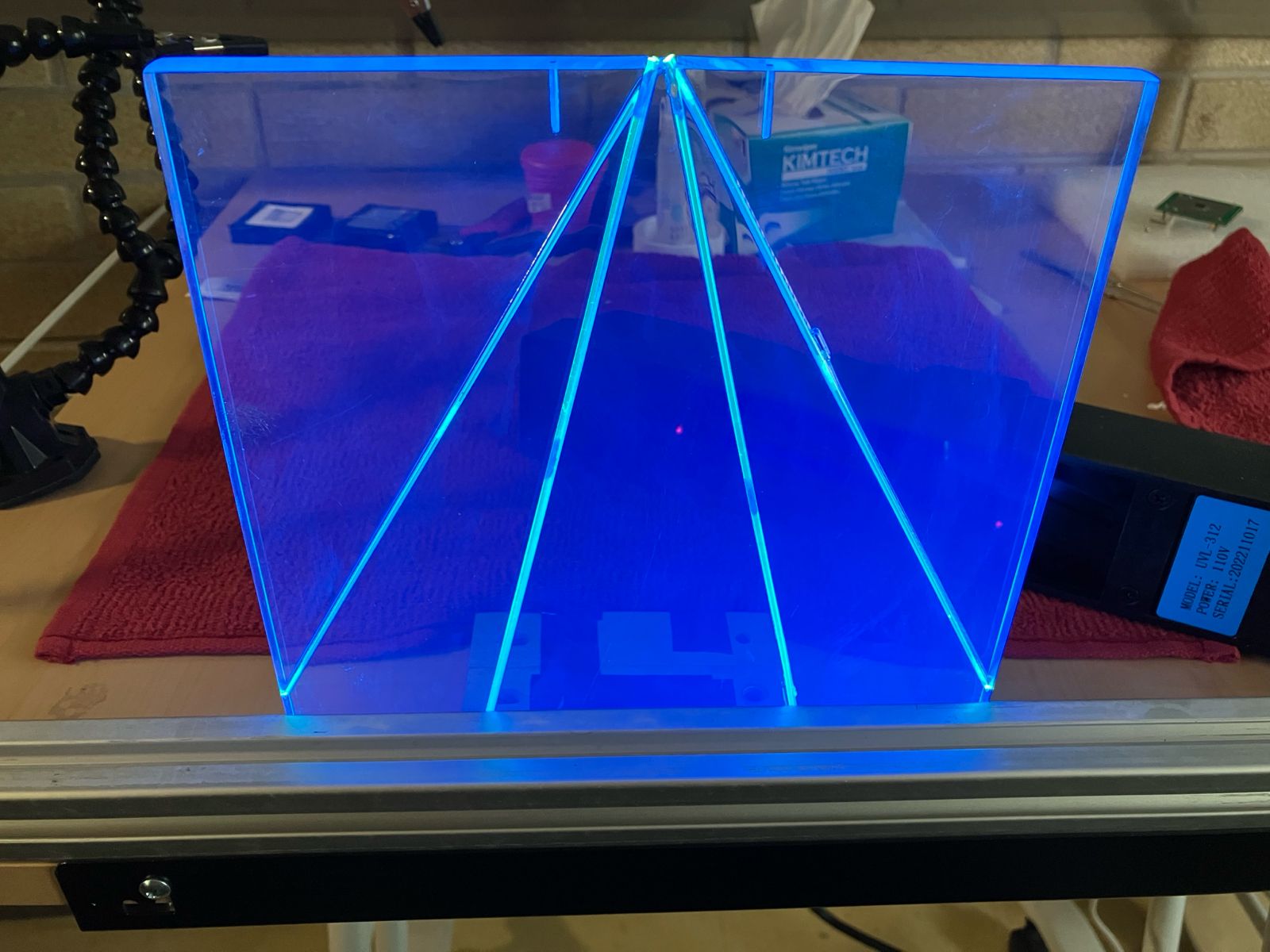}
  \caption{Distribution of the wavelength shifting fibers inside the plastic scintillator. The SiPM is coupled to one of the sides of the scintillator tile.\newline \hfill \newline \hfill} 
  \label{fig-3}
\end{minipage}
\hspace{.05\linewidth}
\begin{minipage}{.45\linewidth}
  \includegraphics[width=\linewidth]{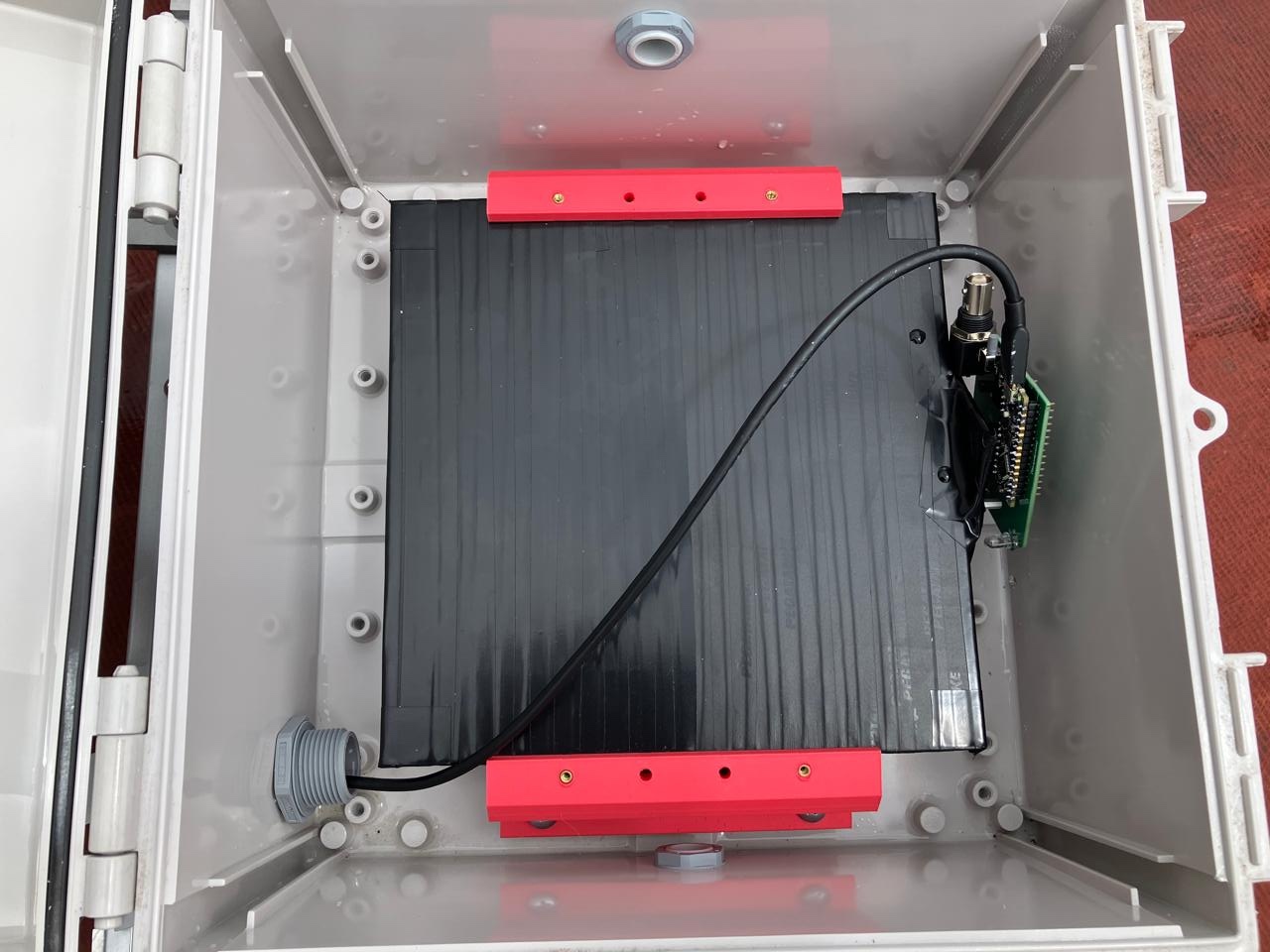}
  \caption{Interior of the electrical enclosures that house the particle detector. The picture shows the mechanical support of the plastic scintillator, as well as the two breathing accessories and the split gland.}
  \label{fig-4}
\end{minipage}
\end{figure}

We use a small array of CosmicWatch (CW) detectors \cite{CW1,CW2} to keep track of the rate of secondary particles. The CW uses an Arduino Nano micro controller to acquire and process the signals produced by SiPMs coupled to plastic scintillators. The uncertainty on the timing of these detectors is of the order of 10 ms \cite{CW3} due to the transfer speed of the serial port. However, the time resolution of the CW is appropriate for our application. The original design of the CW has a small sensitive surface of 25 cm$^{2}$, so we need to increase the area to collect a larger rate in order to be able to observe fluctuations produced by transient phenomena. Our idea is to increase the sensitive surface of each detector by a factor of 16. To do this, we embed wavelength shifting fibers in the scintillators, as shown in Figure \ref{fig-3}. For the current set of detectors, we use Eljen EJ-200 plastic scintillator, but we have also acquired BC408 to build additional detectors. Both models of scintillator have a peak of emission at around 430 nm.

Our scintillators are squares of 20 cm per side, with a width of 1 cm. We use the BCF91AXL wavelength shifting fiber from Luxium, with a diameter of 1.5 mm. This fiber has an emission peak at around 490 nm. By shifting the wavelength, we lose some efficiency in the coupling to the SiPM, but we believe this could be a good compromise in terms of being able to attach a larger scintillator to the electronics. The plastic scintillator is polished and covered with few layers of white Teflon tape. Finally, the scintillator is optically isolated using black tape. The SiPM and the scintillator are coupled with optical gel.

In order to protect the scintillator detectors and the electronics, we use electrical enclosures made of Acrylonitrile Butadiene Styrene. These light gray boxes are perforated to install two breathing accessories, to allow the movement of air inside the box, and a split gland with a sealing ring, for the USB cable that transmits the data to the computer. In order to reduce the temperature of the boxes when exposed to the sun, we partially cover them with Mylar tape. We custom built 3D printed supports for the plastic scintillators. A picture of the interior of the electrical enclosure, the supports and the enclosure accessories are shown in Figure \ref{fig-4}.

We built several prototypes to test different ways to collect the light from the scintillator. The first idea was to direct the light towards the center of the plastic and mount there the electronics as in the original design of the CW. However, we found that, for our larger detectors, it was more efficient to attach the electronics to one side of the scintillator tile. Due to the small sensitive area of the SiPM, 36 mm$^{2}$, we use only four diagonally embedded fibers to collect the light and direct it to the SiPM.

\section{Data acquisition and processing}

The experiment is controlled with an Intel NUC Mini PC. The Data Acquisition System comprises both open-source and proprietary code.  This is because of some of the hardware's reliance on a Windows environment for operation.  Specifically, we use the Cumulus MX \cite{CumulusMX} software to facilitate communication with the weather station, instead of the proprietary software created by Davis. Cumulus MX offers a comprehensive suite for the monitoring of the station.  We store the data in a database at minutes intervals.  For the Boltek mill, we utilize the Boltek ERL-10 Status monitor.  This software logs data in plain text at 0.05-second intervals and archives the information in daily files. For the magnetometer, we utilize Sensys software, specifically the Sensys FGM3D TD Application, which facilitates a continuous stream of messages from the magnetometer. We configure it to read data in intervals of 0.0025 seconds; once this stream is activated, we employ Python \cite{python} to save the recorded data from that day in plain text format. To provide data redundancy, we retain the daily records in a \texttt{TTree} data structure offered by the ROOT framework \cite{root} for subsequent analysis.  The selection of these structures is based on the framework's ability to manage substantial data volumes, ensuring both access and storage efficiency. The data from the camera is processed using the Siril astronomical image processing tool \cite{siril}. The readout of the CWs is done using the Python acquisition code provided by the developers of the CW project.

\section{Preliminary results}

We started making tests of the data acquisition of the scintillators since October of 2024, adding prototypes as they were built to test the detector assembly method. The construction of the mechanical structures for our experiment started in December 2024, and we have been testing and debugging the different data acquisition systems since then. We have been taking data with the four initial prototypes of the scintillator detectors, in the location of the experiment, since May 7, 2025. All of these prototypes have a smaller surface than the design goal size of 20 cm per side.

\begin{figure}[h]
\centering
\begin{minipage}{.45\linewidth}
  \includegraphics[width=\linewidth]{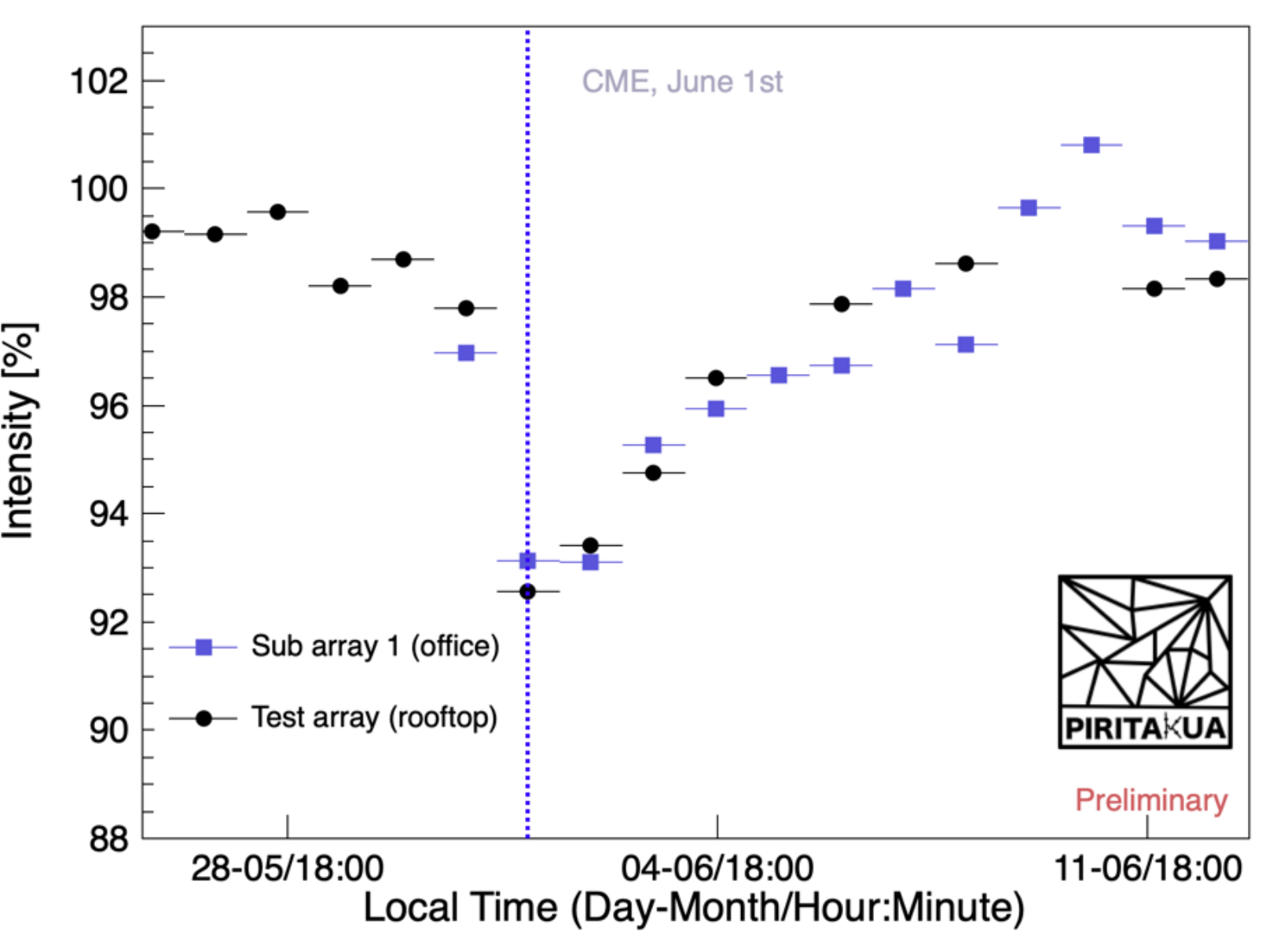}
  \caption{Secondary particle percentage change due to the passage of a CME over Earth. The two markers indicate data from two different testing arrays of scintillators. One sub-array was located in an office space and the other on a rooftop. } 
  \label{fig-5}
\end{minipage}
\hspace{.05\linewidth}
\begin{minipage}{.45\linewidth}
  \includegraphics[width=\linewidth]{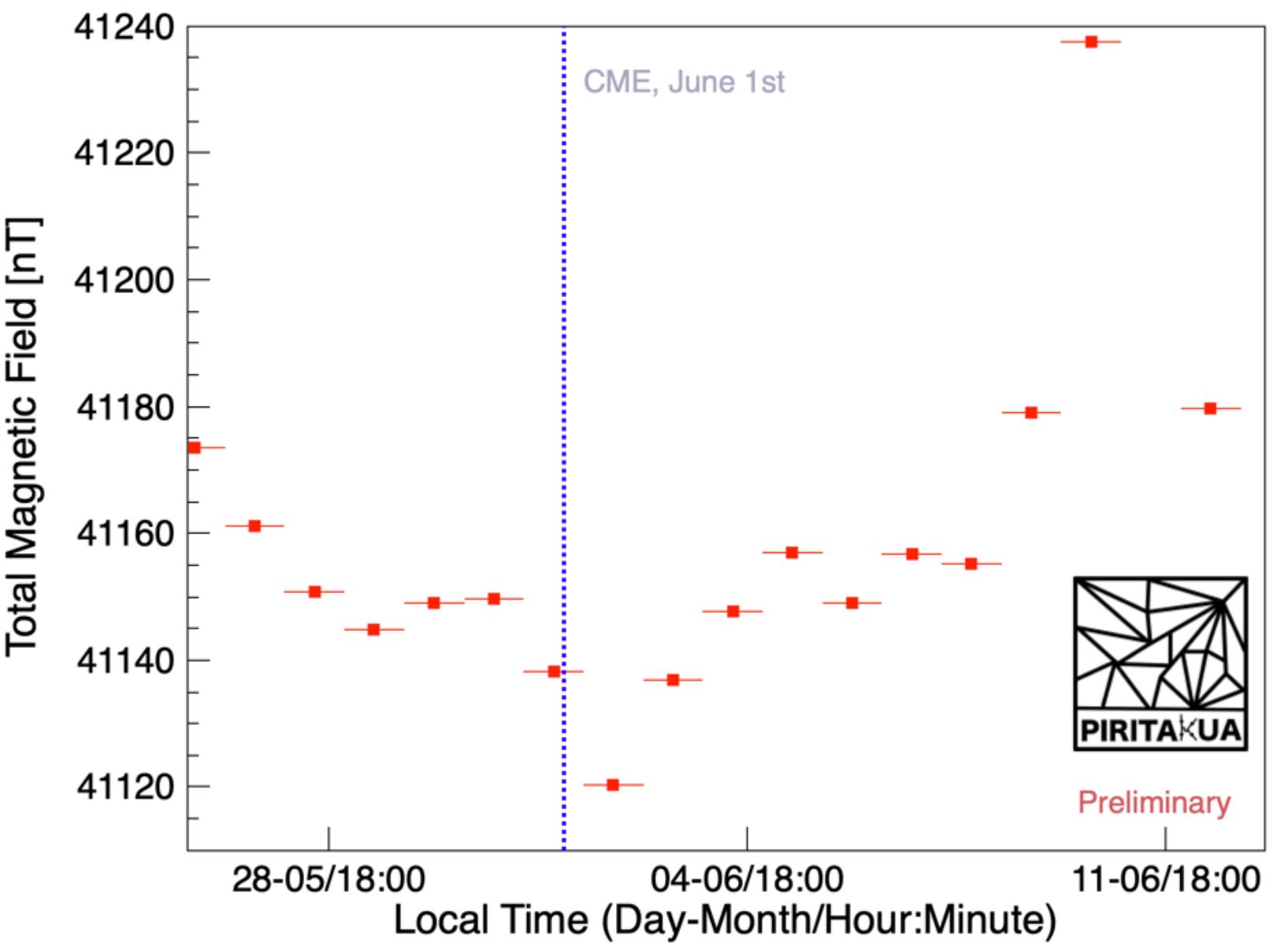}
  \caption{Average local magnetic field, during the same time window as in Figure \ref{fig-5}. The vertical line shows the approximate arrival time of the CME.\newline \hfill}
  \label{fig-6}
\end{minipage}
\end{figure}

At the end of May 2025, a Coronal Mass Ejection (CME) started to pass over Earth. This caused a decrease in the secondary particle rate, and we had enough sensitivity to observe this effect. This is shown in Figure \ref{fig-5}, where the black markers indicate the change in the detection rate intensity as a function of time. We observe a significant decrease of the particle rate, followed by a slow recovery that took approximately one week. By May 30, 2025, we were ready to start testing five additional scintillator detectors, these with the final design dimensions. Before taking the detectors to the rooftop, we test the stability of the data taking in an office space of the same building, located at the ground floor. We were interested in studying the effect of lowering the signal detection threshold used in the configuration of the Arduino. The default value in the Arduino code is of 50 ADC counts ($\approx 15$ mV), with a range from 0 to 1023 ADC counts. We were testing using a reduced value of 30 ADC counts ($\approx 12$ mV) at the time the CME started to pass over Earth.  We decided to keep using this setting with this new subarray of detectors during the duration of the phenomenon. The measurements are shown in Figure \ref{fig-5}. We do not combine the rates of these two sets of detectors since they use different signal thresholds, and the detectors are exposed to different experimental conditions. Moreover, the two independent data sets allow us to confirm the presence of the rate structure associated to the pass of the CME. For comparison, the Mexico City Cosmic Ray Observatory at UNAM\footnote{http://www.cosmicrays.unam.mx/} observed a maximum decrease in the particle rate of $\approx$ 12$\%$, while in our case has a maximum decrease of the order of $\approx$ 8$\%$ in both sets of detectors.

Since we still do not have enough coincident data with both the weather station and the scintillators, we are not ready to perform pressure corrections on the data. Instead, we show the data using daily bins in Figure \ref{fig-5}, so the daily modulation of the rate is absorbed for each bin and makes easier to observe the structure in the rate. Figure \ref{fig-6} shows the local magnetic field values during the same time window as the rate in Figure \ref{fig-5}. The vertical purple line in both Figures shows the approximate time of the arrival of the CME to Earth, on June 1, 2025.  We found that the shape of the magnetic field as a function of time that we observe is similar to the geomagnetic activity reported by the South African National Space Agency\footnote{https://www.sansa.org.za/2025/06/geomagnetic-storm-hits-earth-following-intense-solar-activity/}, with an initial decrease of  before May 28, followed by an additional decrease between June 1 and 2. Soon we will have access to the individual components of the magnetic field, here we show only the sum of the components since are working in the proper alignment of the sensor.

Given the relatively small surface area of our particle detectors, our first goal is to measure TGEs, since these happen during minute long time windows \cite{hhap}. During these phenomena, high-energy electrons emit bremsstrahlung photons that are detectable with ground-level detectors \cite{catalog}. While we add more scintillator area to our array to try to look for ground enhancements, as proof of principle we show an example of the events that we have acquired recently, during the start of the rain season in Mexico City. Our idea is to develop an observation campaign of high-energy phenomena associated to thunderstorms, similar to the work done by the GROWTH collaboration \cite{catalog}.

\begin{figure}[h]
\centering
\includegraphics[width=0.5\linewidth]{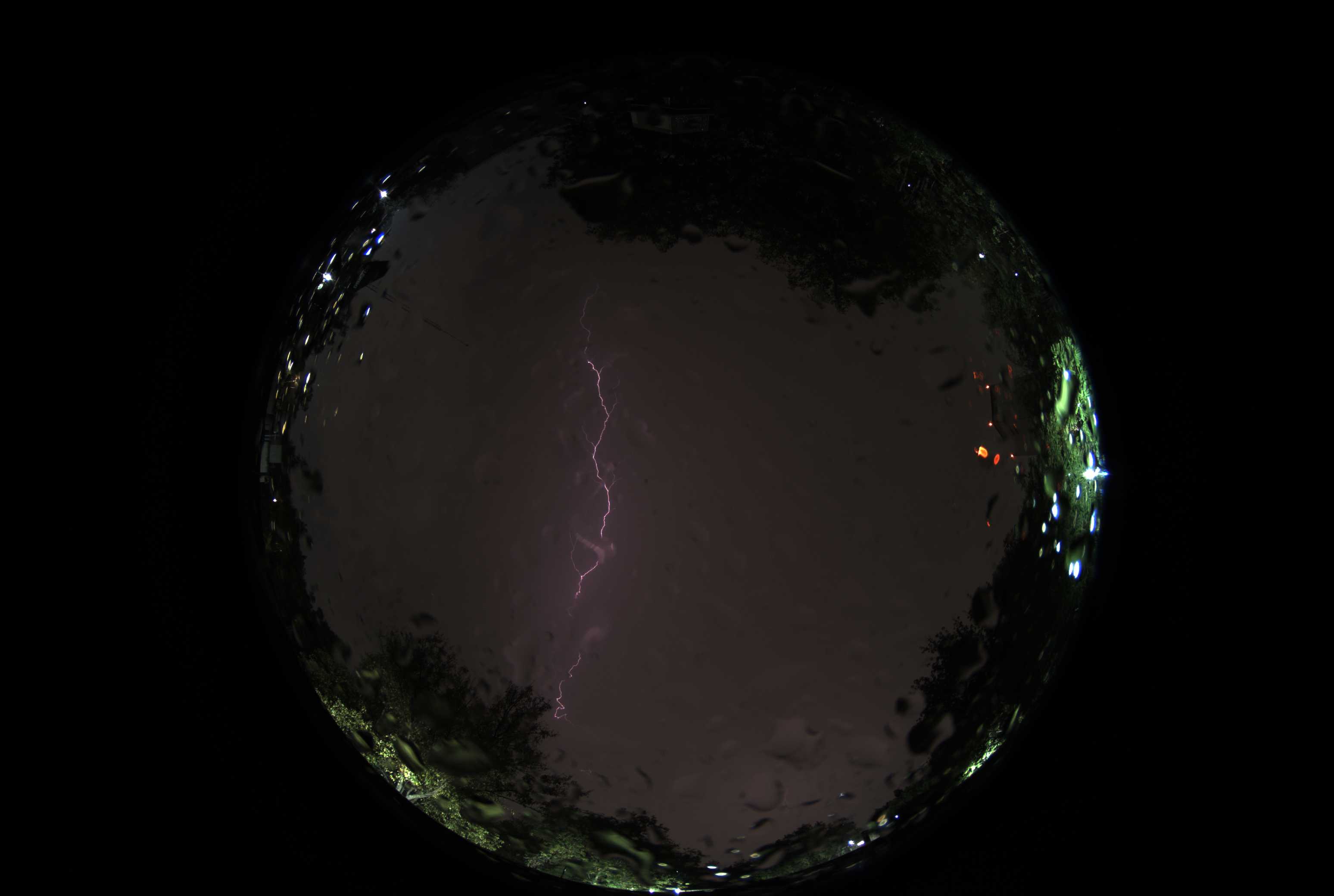}
  \caption{Lightning during the storm of June 11 in Mexico City. The time of this event is shown with red vertical lines in Figures \ref{fig-7} and \ref{fig-8}.}
  \label{fig:lightning}
\end{figure}

Figure \ref{fig:lightning} shows a picture of one of the lightning events that took place during the thunderstorm of June 11 in the south of Mexico City.

\begin{figure}[h]
\centering
\begin{minipage}{.45\linewidth}
  \includegraphics[width=\linewidth]{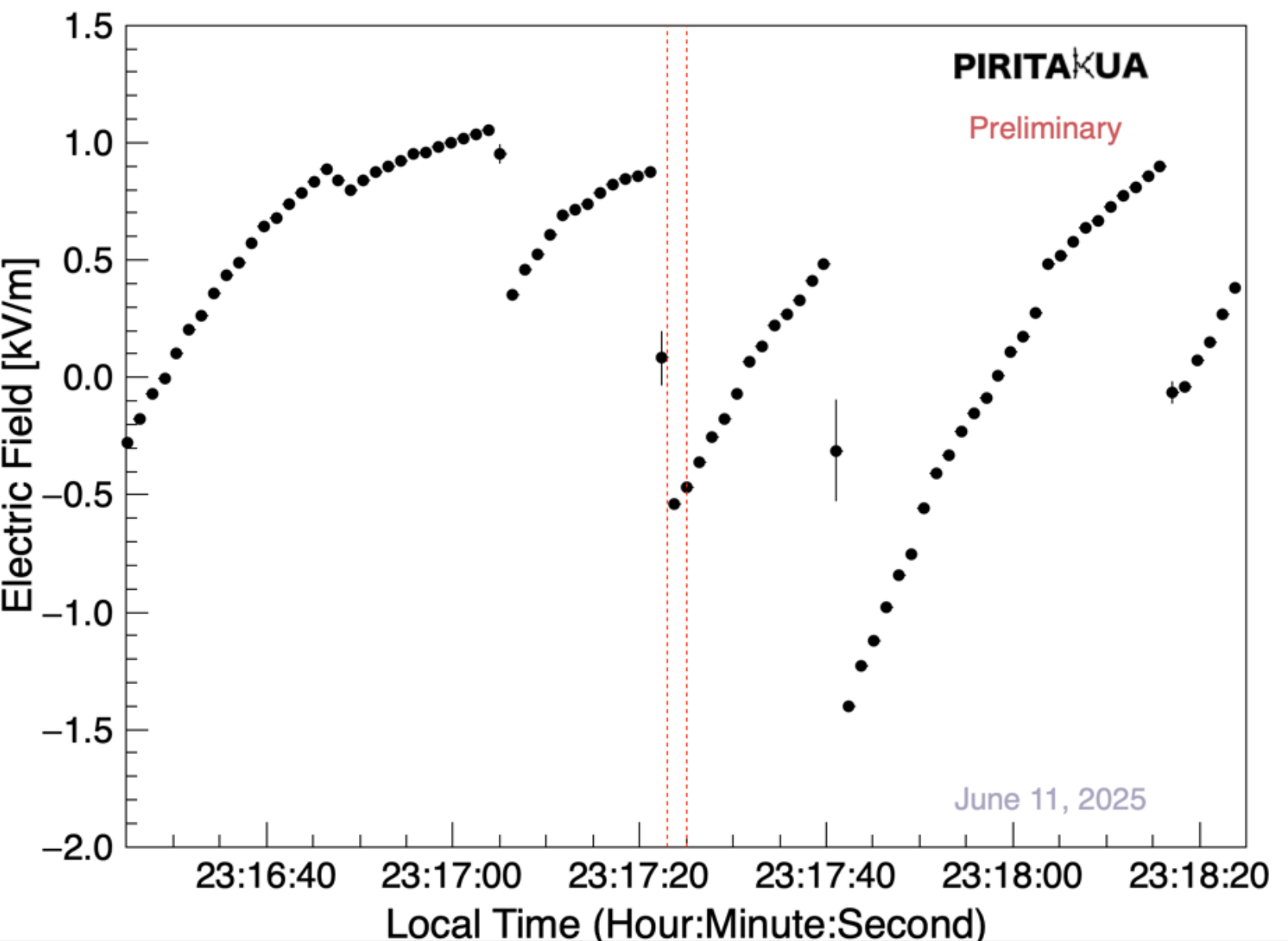}
  \caption{Electric field measurements during a thunderstorm on June 11, 2025. The vertical lines shows the trigger time of a lightning strike captured with the all sky camera, shown in Figure \ref{fig:lightning}.} 
  \label{fig-7}
\end{minipage}
\hspace{.05\linewidth}
\begin{minipage}{.45\linewidth}
  \includegraphics[width=\linewidth]{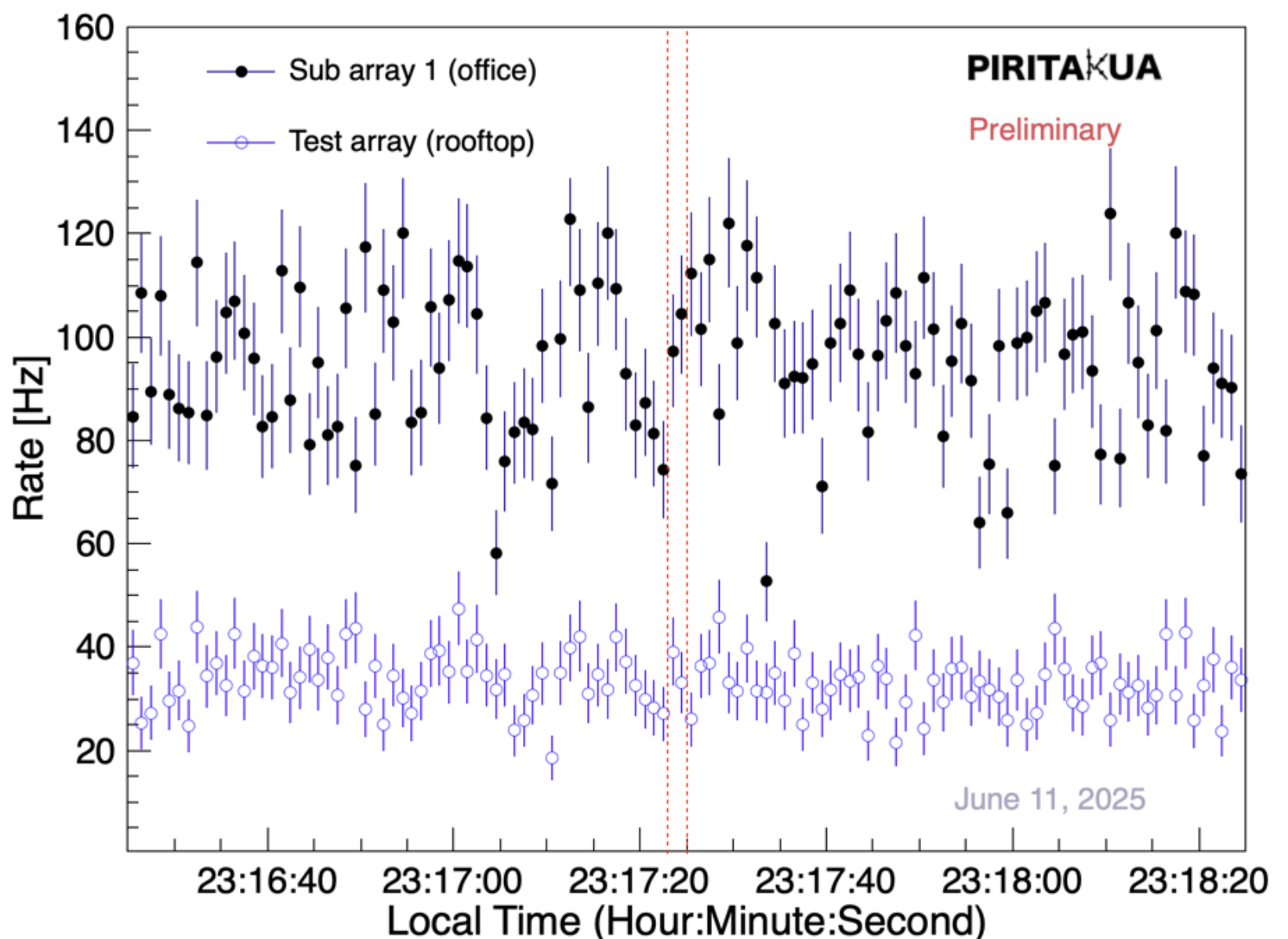}
  \caption{Rate measurements during a thunderstorm on June 11, 2025. The vertical lines shows the trigger time of a lightning strike captured with the all sky camera.\newline \hfill}
  \label{fig-8}
\end{minipage}
\end{figure}

Figure \ref{fig-7} shows the values of the electric field during a time window of one minute around the lightning event shown in Figure \ref{fig:lightning}. The time of the lightning event is shown with red vertical lines. The separation of the vertical lines shows the exposure time of the picture. Figure \ref{fig-8} shows the rate measurements in one-second bins, for the same two-minute time window of Figure \ref{fig-7}, for the two sub arrays of scintillators that were operational at the time. Sub-array 1 refers to the five scintillators that were taking data inside an office space, while the test array refers to four prototype scintillators that were located on the rooftop. These last detectors show a significantly smaller rate mainly due to their smaller physical dimensions, compared to the size of the final version of the detector used in Sub-array 1. It is interesting to notice that there seem to be hints of structures in the detection rate, that possibly correlate with the electric field variations produced by lightning discharges. However, more data is needed before we can confirm that this is what we are observing. Unfortunately, the DAQ of the magnetometer was interrupted during this thunderstorm. We are currently working on a monitoring system to avoid this type of interruptions in the data taking.

\section{Summary}

We have reported on the progress of the construction of a cosmic ray detector at the Physics Institute of UNAM. The goal is to use it to characterize the effect of transient atmospheric phenomena in the production and propagation of secondary particles in air showers. We have shown that even though it is a small-scale experiment, we have been able to observe the effect of a CME in the measured particle rate and hints of the effect of the electric field of thunderstorms in the secondary particle propagation. We anticipate to operate the experiment with 16 scintillator modules by the end of this year, and start to contribute to the study of high-energy atmospheric and cosmic-ray physics.

\acknowledgments

We are grateful to the creators of the CosmicWatch Desktop Muon Detector, for making their design open source. This work was supported by SECIHTI (previously CONACYT), Mexico, grant CF-2023-I-645, and by  UNAM-PAPIIT grant IN102223. This work was supported by Universidad Nacional Aut\'onoma de M\'exico Postdoctoral Program (POSDOC). We thank the support from Maira P\'erez Vielma, Rodrigo Guti\'errez, Fabiola P\'erez Rubio, Alejandra L\'opez Su\'arez, Francisco Marquez, Hesiquio Vargas and Jaime P\'erez.


\begin{thebibliography}{99}

\bibitem{flickering}
Østgaard, N., Mezentsev, A., Marisaldi, M. et al.,
\emph{Flickering gamma-ray flashes, the missing link between gamma glows and TGFs},
\emph{Nature} {\bf 2024} 53–56

\bibitem{lhasso}
C. Yang, X. X. Zhou, H. H. He, D. H. Huang, X. J. Chen, T. Zhou, and K. J. Guo,
\emph{Effect of near-Earth thunderstorm electric fields on the flux
of cosmic ray air showers in LHAASO-KM2A},
\emph{Physical Review D} {\bf 2025} 063023
[{\tt 	arXiv:2410.07925}].

\bibitem{CW1}
Spencer N. Axani,
\emph{The Physics Behind the CosmicWatch Desktop Muon Detectors},
[{\tt 	arXiv:1908.00146}].

\bibitem{CW2}
S.N. Axani, K. Frankiewicz and J.M. Conrad,
\emph{The CosmicWatch Desktop Muon Detector: a
self-contained, pocket sized particle detector},
\emph{JINST} {\bf 2018} P03019

\bibitem{CW3}
S.N. Axani, K. Frankiewicz and J.M. Conrad,
\emph{CosmicWatch: The Desktop Muon Detector. Instruction Manual},
[{\tt 	https://github.com/spenceraxani/CosmicWatch-Desktop-Muon-Detector-v2}].

\bibitem{CumulusMX}
Cumulus MX Development Team, 
\emph{The Cumulus MX weather program},
[{\tt 	https://github.com/cumulusmx/CumulusMX}].

\bibitem{python}
Python Software Foundation, 
\emph{Python 3: An interpreted, interactive, object-oriented programming language},
[{\tt 	https://www.python.org}].

\bibitem{root}
Rene Brun and Fons Rademakers
\emph{ROOT - An Object Oriented Data Analysis Framework},
\emph{Proceedings AIHENP'96 Workshop, Lausanne, Sep. 1996}
\emph{Nucl. Inst. \& Meth. in Phys. Res. A} {\bf 1997} 389, 81-86

\bibitem{siril}
C. Richard et al., 
\emph{Siril: An Advanced Tool for Astronomical Image Processing},
\emph{Journal of Open Source Software}  9 {\bf 2024} 7242

\bibitem{hhap}
Dwyer, J.R., Smith, D.M. \& Cummer, S.A.
\emph{High-Energy Atmospheric Physics: Terrestrial Gamma-Ray Flashes and Related Phenomena},
\emph{Space Sci Rev} {\bf 2012} 173, 133-196

\bibitem{catalog}
Y. Wada, T. Matsumoto, T. Enoto,  K. Nakazawa et al., 
\emph{Catalog of gamma-ray glows during four winter seasons in Japan},
\emph{Phys. Rev. Res.}  3 {\bf 2021} 043117
[{\tt 	arXiv:2108.01829}].

\end{thebibliography}
\end{document}